\documentclass[journal]{IEEEtran}

\usepackage{amsmath,amsfonts}
\usepackage{algorithmic}
\usepackage{array}
\usepackage[caption=false,font=normalsize,labelfont=sf,textfont=sf]{subfig}
\usepackage{textcomp}
\usepackage{stfloats}
\usepackage{url}
\usepackage{verbatim}
\usepackage{graphicx}
\usepackage[utf8]{inputenc}
\usepackage[numbers,sort&compress]{natbib}
\usepackage{booktabs,array,makecell}
\usepackage[ruled,linesnumbered]{algorithm2e}

\usepackage{rotating}
\usepackage{csquotes}
\usepackage{amsmath}
\usepackage{bm}
\usepackage{tabularx}
\usepackage{stfloats}
\usepackage{threeparttable}
\usepackage{url}
\usepackage{soul}
\usepackage{threeparttable}
\soulregister\cite7
\soulregister\ref7
\soulregister\pageref7
\usepackage{amssymb}
\usepackage{soul}
\usepackage{footnote}
\usepackage{colortbl}
\usepackage{soul}
\usepackage{multirow}
\usepackage{pifont}
\usepackage{algorithmic}
\usepackage{booktabs,xcolor,colortbl}
\usepackage{color}
\usepackage{alltt}
\usepackage{hyperref}
\usepackage{enumerate}
\usepackage{siunitx}
\usepackage{epstopdf}
\usepackage{bbding}
\usepackage{pbox}
 \usepackage{amsmath}  
 \usepackage{amssymb}   
 \usepackage{amsthm}
 \usepackage[dvipsnames]{xcolor}

\usepackage{xcolor}

\newcommand{\probP}{\text{I\kern-0.15em P}}
\newcommand{\probE}{\text{I\kern-0.15em E}}

\def\BibTeX{{\rm B\kern-.05em{\sc i\kern-.025em b}\kern-.08em
    T\kern-.1667em\lower.7ex\hbox{E}\kern-.125emX}}
\usepackage{balance}
\begin{document}
\title{Transferable Neuromorphic Protection for Distribution Networks}
\author{Saurabh Prabhakar, \textit{Member, IEEE},  Yubo Song,  \textit{Senior Member, IEEE}, Pascal Schirmer, \textit{Member, IEEE}, Bijaya Ketan Panigrahi, \textit{Fellow, IEEE}, and Subham Sahoo, \textit{Senior Member, IEEE}
}

\maketitle

\begin{abstract}
Protection of converter-dominated distribution networks is challenging due to limited fault currents, operating-mode transitions, bidirectional power flow, and heterogeneous network configurations. This paper proposes a scale-invariant neuromorphic protection framework to enable transferable protection decisions across such networks. The framework integrates asset-based normalization, multi-domain disturbance representation, event-driven spike encoding, leaky integrate-and-fire (LIF) dynamics, adaptive threshold regulation, and directional supervision within a unified mathematical formulation. Local voltage, current, active power, and reactive power measurements are transformed into dimensionless disturbance representations, while latency encoding and membrane integration capture disturbance intensity and temporal evolution. The framework is evaluated across four heterogeneous network configurations: a two-DER inverter-interfaced microgrid, an islanded ring microgrid, the IEEE 13-node feeder, and the CIGRE European medium-voltage benchmark. Across 480 cases, it achieves a median operating time of 7.6~ms, with 94.2\% of cases operating within one power cycle. The results demonstrate selective protection across different electrical ratings and network topologies using a common parameterization without network-specific retuning, highlighting the potential of neuromorphic processing for physically interpretable, communication-free protection of converter-dominated distribution networks.
\end{abstract}


\section{INTRODUCTION}

The growing integration of inverter-based resources (IBRs) is reshaping the protection requirements of modern distribution networks. In contrast to conventional synchronous generators, power-electronic converters generally impose limits on fault-current contribution through their control and protection mechanisms, potentially complicating fault detection under certain operating conditions. The challenge is further compounded by transitions between grid-connected and islanded operation, network reconfiguration, and bidirectional power flow \cite{narang2021overview, hooshyar2017microgrid}. These characteristics can reduce the sensitivity and compromise the coordination of conventional protection schemes in heterogeneous, converter-dominated networks\cite{manson2021practical, quispe2022transmission, aliyu2025comprehensive}. Accordingly, developing protection mechanisms that accommodate variations in equipment ratings, network configurations, and operating conditions without fundamental redesign remains an important research objective \cite{barranco2023adaptive}.

Several approaches have been investigated to address these challenges, including adaptive protection, differential protection, signal-processing-based methods, and time-domain techniques \cite{barranco2023adaptive, dua2022microgrid, liu2022transient, amiri2020integrated, samkari2023time}. For instance, Samkari and Johnson proposed a time-domain protection scheme for microgrids comprising aggregated inverter-based distributed energy resources, demonstrating the potential of time-domain approaches for converter-dominated systems \cite{samkari2023time}. Furthermore, reviews of IBR protection have highlighted the need to assess emerging methods in terms of implementation feasibility, validation maturity, and generalization across diverse power-system conditions \cite{quispe2022transmission, zaben2024machine}. Despite their contributions to specific protection problems, developing a common protection-processing architecture applicable to heterogeneous networks remains an open challenge. In particular, successful validation on one network does not necessarily demonstrate that the underlying decision mechanism can be transferred to systems with different electrical ratings, topologies, and operating conditions \cite{karabulut2025generalization, paladhi2024reliable}.

Neuromorphic computing provides an alternative approach to protection by representing electrical disturbances through event-driven neuronal dynamics rather than relying exclusively on conventional fixed-threshold logic or data-trained classifiers. Leaky integrate-and-fire (LIF) dynamics, in particular, enable disturbance-related inputs to accumulate over time, while spike generation converts the evolving neuronal state into discrete decision events \cite{roy2019towards, tavanaei2019deep}. This provides a means of incorporating both disturbance magnitude and temporal evolution into a physically interpretable protection process. Recent studies have explored neuromorphic principles for microgrid protection, including communication-free operation based on locally processed electrical measurements \cite{prabhakar2026synapse, sahoo2025spike, diao2024neuromorphic, kamoona2026robust}. However, extending these concepts to heterogeneous distribution networks requires more than demonstrating functionality on an individual system. It necessitates a systematic formulation of the complete processing chain and an investigation of its applicability under variations in electrical ratings, network topology, and operating conditions.

An important consideration is that normalization of electrical measurements alone does not establish invariance of the overall protection response. Although per-unit representation provides a common basis for expressing electrical quantities across systems with different nominal ratings, the subsequent disturbance encoding, spike timing, membrane integration, and firing-threshold regulation also determine the final protection decision~\cite{ sahoo2025spike}. Moreover, the electrical dynamics of inverter-interfaced distributed energy resources (DERs), arising from energy-storage and resistive elements, can be represented through an equivalent RC-based leaky integrate-and-fire (LIF) model, providing a physical foundation for neuromorphic processing~\cite{prabhakar2026synapse}. Thus, transferability depends on the consistent behaviour of the complete processing chain rather than on normalization in isolation. This raises the question of whether the neuromorphic protection mechanism can maintain effective operation across heterogeneous distribution networks without network-specific modification of its underlying protection logic.

To address this challenge, this paper proposes a scale-invariant neuromorphic protection framework that integrates asset-based normalization, multi-domain disturbance representation, event-driven spike encoding, LIF dynamics, and adaptive threshold regulation. Locally measured voltage, current, active power, and reactive power are normalized using system-specific bases and processed through neuronal dynamics, while directional supervision ensures selective fault tripping. The central objective is to assess whether a common protection parameterization can maintain effective protection across networks with different electrical ratings, topologies, and operating modes without modifying the underlying decision logic. The framework is evaluated on radial and ring microgrids, an islanded microgrid, the IEEE 13-node feeder, and the CIGRE medium-voltage benchmark.

The main contributions are summarized as follows:

\begin{enumerate}
    \item \textit{Unified scale-invariant protection formulation:} An integrated mathematical framework combining asset-based normalization, multi-domain disturbance representation, event-driven encoding, LIF dynamics, and adaptive threshold regulation is developed for heterogeneous distribution networks.

    \item \textit{Cross-network transferability:} The ability of a common protection parameterization to maintain consistent protection response, operating time, and selectivity across different electrical ratings, topologies, and operating modes is systematically assessed.

    \item \textit{Comprehensive validation:} The proposed framework is validated across radial and ring microgrids, an islanded microgrid, the IEEE 13-node feeder, and the CIGRE medium-voltage benchmark.
\end{enumerate}

The remainder of this paper is organized as follows. Section~II presents the mathematical formulation. Section~III describes the test systems and protection performance. Section~IV analyses operating-time characteristics, protection sensitivity, and comparisons with existing protection schemes. Section~V concludes the paper and discusses future research directions.

\begin{figure*}[t]
  \centering
  \includegraphics[width=0.9\textwidth]{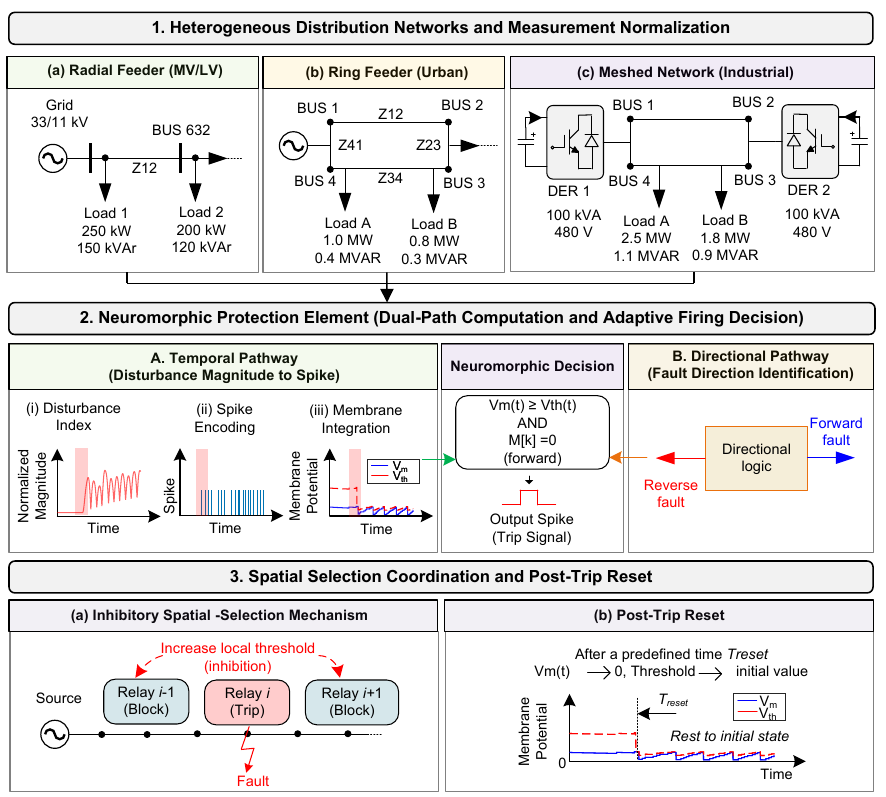}
  \vspace{-12pt}
  \caption{Overall operational flow of the proposed neuromorphic protection framework, illustrating asset-based normalization, multi-domain disturbance formulation, event-driven spike encoding, leaky integrate-and-fire dynamics, adaptive firing-threshold regulation, directional supervision, and the final protection decision.}
  \label{fig1}
\end{figure*}

\section{Mathematical Formulation of the Proposed Framework}

The proposed framework comprises three integrated stages, as illustrated in Fig.~\ref{fig1}. The first stage establishes a common measurement representation across heterogeneous distribution networks through asset-based normalization. The second implements the neuromorphic protection element, in which disturbance magnitude drives event-based spike encoding and neuronal integration, while an independent directional pathway supervises the firing decision. The third coordinates spatial selectivity among neighbouring relays and incorporates post-trip membrane reset. The following subsections formulate these stages and establish the conditions for consistent protection behaviour across different network configurations.

\subsection{System Representation and Asset-Based Normalization}

The considered distribution network comprises inverter-interfaced distributed energy resources (DERs), distribution lines, and buses operating under different topologies and operating modes. As illustrated in Fig.~\ref{fig1}, the proposed framework accommodates heterogeneous radial, ring, and meshed configurations. At protection terminal $r$, the locally measured three-phase voltage and current are denoted by $\mathbf{v}_r[k]$ and $\mathbf{i}_r[k]$, respectively, while $P_r[k]$ and $Q_r[k]$ represent the corresponding active and reactive powers. The local measurement vector is defined as
\begin{equation}
\mathbf{x}_r[k]=
\begin{bmatrix}
\mathbf{v}_r^{T}[k] &
\mathbf{i}_r^{T}[k] &
P_r[k] &
Q_r[k]
\end{bmatrix}^{T}.
\label{eq:measurement_vector}
\end{equation}

To accommodate variations in system ratings, the measured quantities are normalized using their respective voltage, current, and power bases. For a three-phase system, the current and impedance bases are
\begin{equation}
I_{\mathrm{base},r}=
\frac{S_{\mathrm{base},r}}
{\sqrt{3}V_{\mathrm{base},r}},
\qquad
Z_{\mathrm{base},r}=
\frac{V_{\mathrm{base},r}^{2}}
{S_{\mathrm{base},r}},
\label{eq:system_bases}
\end{equation}
where $V_{\mathrm{base},r}$ is the nominal line-to-line voltage and $S_{\mathrm{base},r}$ is the three-phase apparent-power base. The corresponding per-unit voltage and current are
\begin{equation}
\mathbf{v}_{r,\mathrm{pu}}[k]=
\frac{\mathbf{v}_r[k]}{V_{\mathrm{base},r}},
\qquad
\mathbf{i}_{r,\mathrm{pu}}[k]=
\frac{\mathbf{i}_r[k]}{I_{\mathrm{base},r}},
\label{eq:normalized_vi}
\end{equation}
while the normalized active and reactive powers are
\begin{equation}
P_{r,\mathrm{pu}}[k]=
\frac{P_r[k]}{S_{\mathrm{base},r}},
\qquad
Q_{r,\mathrm{pu}}[k]=
\frac{Q_r[k]}{S_{\mathrm{base},r}}.
\label{eq:normalized_pq}
\end{equation}

The deviations from the local prefault operating condition are represented by the normalized disturbance vector:
\begin{equation}
\Delta\mathbf{x}_{r,\mathrm{pu}}[k]
=
\mathbf{x}_{r,\mathrm{pu}}[k]
-
\mathbf{x}_{r,\mathrm{pu}}^{0},
\label{eq:normalized_deviation}
\end{equation}
where $\mathbf{x}_{r,\mathrm{pu}}^{0}$ denotes the corresponding prefault reference vector. This representation expresses local electrical variations relative to the equipment ratings and prefault operating point, thereby reducing dependence on absolute system magnitudes.

The scaling property of per-unit normalization is established by considering a physical quantity $x(t)$ and its corresponding base $X_{\mathrm{base}}$. Under a common positive scaling factor $c$, they transform as
\begin{equation}
x'(t)=cx(t),
\qquad
X'_{\mathrm{base}}=cX_{\mathrm{base}}.
\label{eq:scaling_transformation}
\end{equation}
The resulting per-unit quantity therefore remains invariant:
\begin{equation}
x'_{\mathrm{pu}}(t)
=
\frac{x'(t)}{X'_{\mathrm{base}}}
=
\frac{cx(t)}{cX_{\mathrm{base}}}
=
\frac{x(t)}{X_{\mathrm{base}}}
=
x_{\mathrm{pu}}(t).
\label{eq:scaling_invariance}
\end{equation}

Thus, per-unit normalization preserves the measurement representation when physical quantities and their corresponding bases are scaled consistently. However, scale invariance of the complete protection response additionally requires consistency in disturbance formulation, spike encoding, membrane dynamics, and firing-threshold regulation, as developed in the following subsections.

\subsection{Multi-Domain Disturbance Formulation and Event-Driven Spike Encoding}

The normalized deviations in (5) are aggregated into a scalar disturbance index:
\begin{equation}
D_r(t) =
|\Delta V_{r,\mathrm{pu}}(t)|
+
|\Delta I_{r,\mathrm{pu}}(t)|
+
|\Delta P_{r,\mathrm{pu}}(t)|
+
|\Delta Q_{r,\mathrm{pu}}(t)|.
\tag{8}
\end{equation}

Here, voltage and current deviations represent scalar features extracted from local measurements. The index captures aggregate disturbance magnitude without empirically fitted weights, while directional and sequence-domain information is retained for subsequent protection discrimination.

Event-driven encoding is activated when $D_r(t)$ exceeds the predefined threshold:
\begin{equation}
\mathcal{E}_r(t) =
\begin{cases}
1, & D_r(t) \geq D_{\mathrm{th}},\\
0, & D_r(t) < D_{\mathrm{th}}.
\end{cases}
\tag{9}
\end{equation}

For activated events, latency-based encoding maps disturbance magnitude to an event latency:
\begin{equation}
\tau_r(t) =
\operatorname{clip}
\left(
\frac{K_{\tau}}{D_r(t)},
\tau_{\min},
\tau_{\max}
\right),
\qquad D_r(t) \geq D_{\mathrm{th}},
\tag{10}
\end{equation}
where $K_{\tau}$ is the latency-scaling coefficient, and $\tau_{\min}$ and $\tau_{\max}$ are the latency bounds. Thus, larger disturbances produce shorter latencies. The encoded events drive subsequent neuronal integration, while subthreshold disturbances remain unencoded.

\subsection{Leaky Integrate-and-Fire Dynamics and Protection Decision}

The encoded disturbance drives neuronal integration, while an independent directional pathway identifies fault direction from local sequence-domain measurements. The normalized directional index is defined as
\begin{equation}
\Gamma_r[k]=
\frac{
\operatorname{Re}\left\{
\Delta V_{1,r}[k]\Delta I_{1,r}^{*}[k]
\right\}
}{
|\Delta V_{1,r}[k]|\,
|\Delta I_{1,r}[k]|+\epsilon
},
\tag{11}
\end{equation}
\noindent where $\Delta V_{1,r}[k]$ and $\Delta I_{1,r}[k]$ are the incremental positive-sequence voltage and current at relay $r$, $(\cdot)^*$ denotes complex conjugation, $\operatorname{Re}\{\cdot\}$ extracts the real component, and $\epsilon$ prevents division by zero. The binary directional state is determined using the forward-polarity reference:
\begin{equation}
M[k]=
\begin{cases}
0, & s_{\mathrm{fwd}}\Gamma_r[k]>\gamma_{\mathrm{dir}}
\ \land\ D_r[k]>D_{\mathrm{th}},\\
1, & s_{\mathrm{fwd}}\Gamma_r[k]<-\gamma_{\mathrm{dir}}
\ \land\ D_r[k]>D_{\mathrm{th}},\\
M[k-1], & \text{otherwise},
\end{cases}
\tag{12}
\end{equation}
\noindent where $s_{\mathrm{fwd}}\in\{-1,+1\}$ defines the forward polarity and $\gamma_{\mathrm{dir}}$ is the directional threshold. Here, $M[k]=0$ and $M[k]=1$ indicate forward and reverse directions, respectively, with the previous state retained when directional evidence is inconclusive.

The encoded input is integrated through a leaky integrate-and-fire (LIF) neuron, whose membrane potential evolves as
\begin{equation}
\tau_m\frac{dV_m(t)}{dt}
=
-\left[V_m(t)-V_{\mathrm{rest}}\right]
+
R_m I_{\mathrm{syn}}(t),
\tag{13}
\end{equation}
\noindent where $V_m(t)$ and $V_{\mathrm{rest}}$ denote the membrane and resting potentials, $\tau_m$ is the membrane time constant, $R_m$ is the membrane resistance, and $I_{\mathrm{syn}}(t)$ represents the encoded synaptic input. This dynamics accumulates successive excitations while attenuating isolated disturbances through membrane leakage.

Neuronal firing occurs when the membrane potential reaches the effective threshold:
\begin{equation}
s_r[k]=
\begin{cases}
1, & V_m[k]\geq\Theta_r[k],\\
0, & V_m[k]<\Theta_r[k],
\end{cases}
\tag{14}
\end{equation}
\noindent where $s_r[k]$ denotes the neuronal output of relay $r$. The threshold is adaptively regulated according to disturbance magnitude and direction:
\begin{equation}
\Theta_r[k]=
\begin{cases}
\Theta_0+V_m[k-1], & M[k]=1,\\[4pt]
\Theta_0 e^{-\lambda_{\mathrm{fast}}D_r[k]/D_{\mathrm{th}}}
+\dfrac{1}{2}V_m[k-1], & M[k]=0.
\end{cases}
\tag{15}
\end{equation}
\noindent Here, $\Theta_0$ is the nominal threshold, $D_r[k]$ and $D_{\mathrm{th}}$ denote the normalized disturbance index and activation threshold, respectively, and $\lambda_{\mathrm{fast}}$ controls threshold sensitivity. The forward-direction threshold decreases with increasing disturbance magnitude, facilitating firing, whereas the higher reverse-direction threshold restrains unintended operation. Following firing, the membrane undergoes the prescribed reset, with an optional refractory interval $T_{\mathrm{ref}}$.

The trip command is issued only when neuronal firing coincides with forward-direction permission:
\begin{equation}
\operatorname{Trip}_r[k]=
\begin{cases}
1, & s_r[k]=1\ \land\ M[k]=0,\\
0, & \text{otherwise}.
\end{cases}
\tag{16}
\end{equation}
Thus, neuronal integration establishes the disturbance response, while directional supervision determines trip eligibility. Algorithm~\ref{alg1} summarizes this processing sequence, including the automatic determination of normalization bases from local ratings $(V_{LL},S)$, with the decision parameters in Table~I retained across the evaluated networks.

\begin{figure*}[t]
  \centering
  \includegraphics[width=\textwidth]{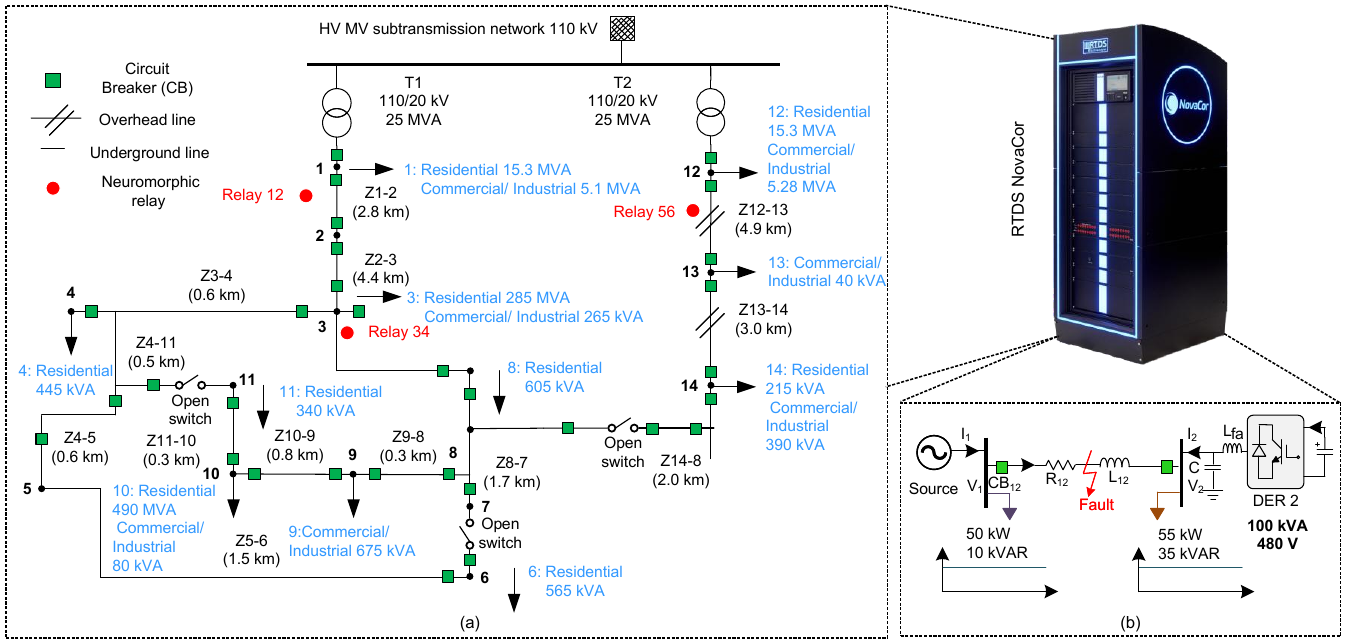}
  \vspace{-6pt}
  \caption{Test networks for Paper 2 validation: (a) the CIGRE European medium-voltage benchmark network, used to assess operation on a larger meshed system. and (b) a two-DER inverter-interfaced microgrid }
  \label{fig2}
\end{figure*}

\subsection{Scale-Invariance Conditions and Cross-Network Transferability}

The proposed framework achieves cross-network transferability through a consistent normalized disturbance representation and common neuromorphic parameterization. The complete protection mechanism is expressed as
\begin{equation}
\operatorname{Trip}_r(t)=
\mathcal{F}_{\boldsymbol{\theta}}
\left(
\Delta\mathbf{x}_{r,\mathrm{pu}}(t)
\right),
\label{eq:parameterized_protection}
\end{equation}
where $\mathcal{F}_{\boldsymbol{\theta}}(\cdot)$ represents the complete protection chain, including disturbance formulation, spike encoding, membrane integration, threshold regulation, and trip generation, governed by the common parameter vector $\boldsymbol{\theta}$.

For two networks, $a$ and $b$, with different electrical ratings but identical normalized disturbance trajectories, strict decision invariance requires
\begin{equation}
\operatorname{Trip}_a(t)=
\operatorname{Trip}_b(t).
\tag{16}
\end{equation}
Identical operating responses additionally require equal trip times:
\begin{equation}
t_{\mathrm{trip},a}=
t_{\mathrm{trip},b},
\tag{17}
\end{equation}
provided that the sampling intervals, encoding parameters, and neuronal dynamics are identical. Thus, equivalent trip decisions do not necessarily imply identical operating times.

This equivalence requires consistent latency mapping and limiting values in (10), together with common membrane and threshold parameters. The discrete-time processing chain is represented as
\begin{equation}
\begin{aligned}
\mathbf{z}_{k+1}
&=
\mathcal{G}\left(
\mathbf{z}_k,\mathbf{u}_k;
\boldsymbol{\theta}
\right),\\
\operatorname{Trip}[k]
&=
\mathcal{H}\left(
\mathbf{z}_k;
\boldsymbol{\theta}
\right),
\end{aligned}
\tag{18}
\end{equation}
where $\mathbf{u}_k$ and $\mathbf{z}_k$ denote the encoded input and internal neuromorphic states, respectively. Identical input sequences, equivalent initial states, and identical state-update, decision, and reset functions ensure equivalent states and trip outputs by induction.

These conditions establish the basis for transferring a common protection architecture across heterogeneous networks. However, practical transferability does not require identical normalized disturbance trajectories. Variations in operating conditions, network impedance, converter characteristics, fault location, and equipment parameters produce network-dependent disturbance signatures despite asset-based normalization. Different sampling rates can also be accommodated through consistent temporal alignment and appropriate adjustment of sampling-dependent parameters. Accordingly, cross-network evaluation assesses whether the common parameterization maintains effective protection performance under these variations.

\subsection{Conditional Scale Invariance and Cross-Network Transferability}

Conditional scale invariance requires equivalent relay responses under identical normalized input trajectories, initial states, and implementation parameters. In contrast, practical transferability concerns dependable and secure operation across different disturbance trajectories using a common parameterization. The cross-network evaluations examine the latter property.

The final protection decision combines neuronal firing with directional supervision. At sampling instant $k$, neuronal firing is determined by
\begin{equation}
F[k]=
\begin{cases}
1, & V_m[k]\geq\Theta[k],\\
0, & \text{otherwise},
\end{cases}
\tag{19}
\end{equation}
\noindent where $V_m[k]$, $\Theta[k]$, and $F[k]$ denote the membrane potential, firing threshold, and neuronal firing state, respectively. Directional selectivity is incorporated through $M[k]\in\{0,1\}$, where $M[k]=0$ and $M[k]=1$ represent forward and reverse directions, respectively. Accordingly, tripping is permitted only when neuronal firing coincides with the forward-direction condition:
\begin{equation}
\operatorname{Trip}[k]=
\begin{cases}
1, & F[k]=1\ \land\ M[k]=0,\\
0, & \text{otherwise}.
\end{cases}
\tag{20}
\end{equation}
\noindent The directional state is obtained from the local discrimination mechanism described in the preceding subsection. Positive-sequence admittance information supports disturbance assessment and directional classification but does not independently initiate tripping unless included in the implemented relay logic.

Following a trip, the membrane potential undergoes a partial reset to prevent immediate retriggering while retaining residual excitation:
\begin{equation}
V_m[k^+]=
\mathcal{R}\left(
V_m[k^-],\Theta[k]
\right),
\tag{21}
\end{equation}
\noindent where $V_m[k^-]$ and $V_m[k^+]$ denote the membrane potentials immediately before and after reset, respectively, and $\mathcal{R}(\cdot)$ represents the adopted reset function.

The combined disturbance encoding, neuronal integration, adaptive threshold regulation, directional supervision, and reset establish the protection mechanism evaluated across radial and ring microgrids, the IEEE 13-node feeder, and the CIGRE medium-voltage benchmark in terms of protection response, operating time, and selectivity.

\begin{algorithm}[!t]
\footnotesize
\DontPrintSemicolon
\caption{Self-Tuning Neuromorphic Protection Core}
\label{alg1}
\KwIn{nameplate $V_{LL},S$; local $V,I,P,Q$, sequence $V_1,I_1,I_2,I_0$;
      references $V_0,I_0^{*},P_0,Q_0$; membrane $m$}
\KwOut{trip spike $s$; first-to-spike time $t_\mathrm{F}$}
\BlankLine
\textbf{1) Per-unit bases from ratings}\;
$V_b=V_{LL}/\sqrt3,\ \ I_b=S/(\sqrt3V_{LL}),\ \ Y_b=I_b/V_b$\;
\BlankLine
\textbf{2) Disturbance index (unit weights)}\;
$D=\sum_{X\in\{V,I,P,Q\}}|X-X_0|/X_b$\;
\If{$D>D_{th}$, $t\ge t_{act}$, $|\dot D|>\eta$}{
  emit spike at latency $\tau=\max\!\big(\tfrac{1}{1+kD},\,2\Delta t\big)$\;
}
\BlankLine
\textbf{3) Directional lock} $L\in\{0,1\}$\;
$\Psi_A=I_2/V_1,\ \ \Psi_\Delta=|\Delta I_1|/|\Delta V_1|$\;
\BlankLine
\textbf{4) Adaptive threshold} $\Theta$\;
\uIf{$L=1$ \textup{(external)}}{$\Theta=U_r+0.5m$}
\uElseIf{fault confirmed}{$\Theta=U_r\,e^{-\lambda D/U_0}+0.5m$}
\Else{$\Theta=U_r/(1+(\Psi_A/\Psi_c)^2)+\Gamma+1.2m$}
\BlankLine
\textbf{5) Integrate-and-fire}\;
$m\gets m\,e^{-\Delta t/\tau_{lk}}$;\ \ if spike: $m\mathrel{+}{=}(\Delta t/\tau_m)R$\;
\If{$m\ge\Theta$}{
  $s\gets1$;\ reset $m\gets m-\Theta$;\ latch $t_\mathrm{F}$\;
}
\end{algorithm}

\begin{table}[!t]
    \renewcommand{\arraystretch}{1.2}
    \setlength{\tabcolsep}{3pt}
    \caption{Benchmark Networks for Cross-Network Validation}
    \label{tab1}
    \centering
    \scriptsize
    \begin{tabular}{@{}l cccc@{}}
        \toprule
        \textbf{Parameter} &
        \makecell{\textbf{Two-DER}\\\textbf{$\mu$grid}} &
        \makecell{\textbf{Islanded}\\\textbf{ring}} &
        \makecell{\textbf{IEEE}\\\textbf{13-node}} &
        \makecell{\textbf{CIGRE}\\\textbf{MV}}\\
        \midrule
        \makecell[l]{Nominal\\voltage (L--L)} & 480 V & 480 V & 4.16 kV & 20 kV\\
        Frequency & 60 Hz & 60 Hz & 60 Hz & 50 Hz\\
        Base power/rating & \makecell{100\\kVA/DER} & \makecell{100\\kVA/DER} & \makecell{Benchmark\\feeder} & 25 MVA\\
        Topology & \makecell{Parallel--\\radial} & \makecell{Closed-\\loop ring} & \makecell{Mixed\\ring--radial} & \makecell{Meshed\\multi-feeder}\\
        Operating mode & \makecell{Grid-\\connected} & Islanded & \makecell{Grid-\\connected} & \makecell{Grid-\\connected}\\
        \midrule
        \multicolumn{5}{@{}l@{}}{\textit{Shared constants:}
        $k=400,\quad U_r=1.0,\quad \lambda=0.5$}\\
        \bottomrule
    \end{tabular}
\end{table}

\section{Performance Evaluation}

\subsection{Real-Time Test Systems and Validation Scenarios}

The proposed framework is evaluated across heterogeneous distribution networks to examine its protection performance under different electrical ratings, network topologies, and operating modes. The selected test systems comprise two-DER inverter-interfaced and ring microgrids, an islanded microgrid, the IEEE 13-node distribution feeder, and the CIGRE European medium-voltage benchmark. The network configurations and key characteristics are presented in Fig.~\ref{fig2} and Table~\ref{tab1}, respectively. Real-time digital simulation (RTDS) is further employed to assess the execution of the proposed protection algorithm under real-time operating conditions.

The nominal frequencies of the considered systems are 50~Hz for the CIGRE medium-voltage benchmark and 60~Hz for the remaining test systems. Accordingly, one fundamental cycle corresponds to 20~ms at 50~Hz and 16.67~ms at 60~Hz. All cycle-based protection operating times are referenced to the nominal frequency of the corresponding test system.

The two-DER microgrid provides a test environment for evaluating protection responses under inverter-interfaced operating conditions, while the CIGRE benchmark extends the assessment to a larger meshed distribution network. The remaining configurations enable evaluation across different network topologies and operating modes using a common protection formulation.

\subsection{Protection Performance in a Two-DER Inverter-Interfaced Microgrid}

The two-DER inverter-interfaced microgrid in Fig.~\ref{fig2}(b) is subjected to two internal-fault scenarios: an A-phase-to-ground (AG) fault and a BCG fault, both initiated at $t_f=0.0530~\mathrm{s}$.

\begin{figure}[!htb]
\centering
\hspace*{-0.8cm}
\includegraphics[scale=1.1]{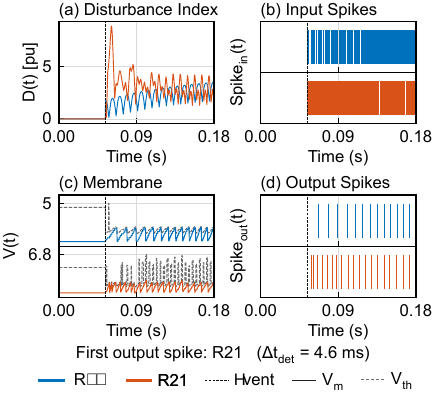}
\caption{Temporal protection signatures for an AG fault in the two-DER inverter-interfaced microgrid---(a) disturbance index $D[k]$, (b) latency-coded input spikes, (c) membrane potential $V_m$ against the adaptive threshold $V_{\mathrm{th}}$, and (d) output spikes. Both relays R12 and R21 operate, the first firing $\Delta t_{\mathrm{det}}$ after disturbance onset.}
\label{fig3}
\end{figure}

\begin{figure}[!htb]
\centering
\hspace*{-0.8cm}
\includegraphics[scale=1.1]{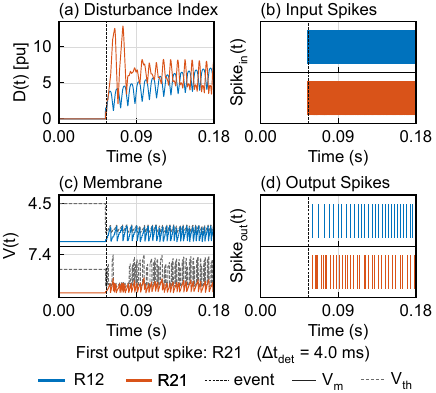}
\caption{Temporal protection signatures for a BCG fault in the two-DER microgrid – following the same pipeline, both relays R12 and R21 correctly operate within one cycle.}
\label{fig4}
\end{figure}

For the AG fault, Fig.~\ref{fig3} illustrates the disturbance index, latency-coded input spikes, membrane dynamics, and output spikes. Both relays $R_{12}$ and $R_{21}$ operate, with $R_{21}$ generating the first output spike after 4.6~ms. Similarly, under the BCG fault, Fig.~\ref{fig4} shows that both relays respond within one cycle, with $R_{21}$ again generating the first output spike, after 4.0~ms.

The detection intervals of 4.6 and 4.0~ms demonstrate rapid responses to both fault conditions. The consistent first response from $R_{21}$, despite different fault characteristics, reflects the influence of local disturbance signatures on neuronal integration and spike generation. The operation of both relays further demonstrates the coordinated protection response under the two evaluated internal faults.

\begin{figure}[!htb]
\centering
\hspace*{-0.8cm}
\includegraphics[scale=1.1]{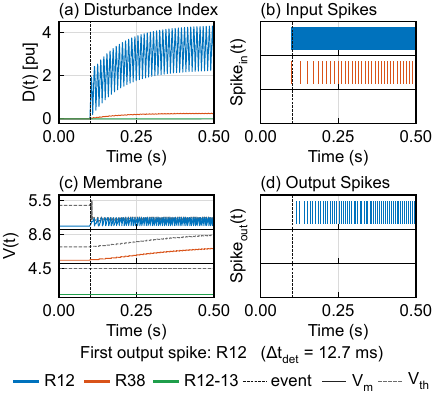}
\caption{Temporal protection signatures for an AG fault on Line 12 of the CIGRE MV network – (a) disturbance index, (b) input spikes, (c) membrane dynamics, and (d) output spikes. Only the faulted-line relay R12 operates; R38 and R12-13 are restrained.}
\label{fig5}
\end{figure}

\begin{figure}[!htb]
\centering
\hspace*{-0.8cm}
\includegraphics[scale=1.1]{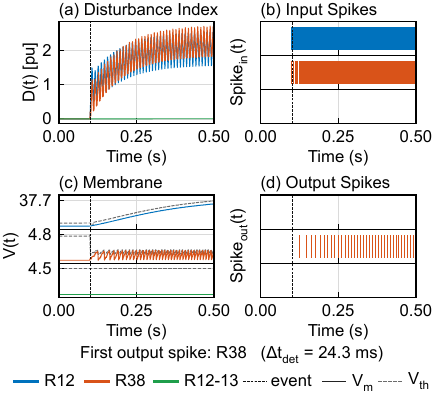}
\caption{Temporal signatures for a BCG fault on Line 38 (CIGRE MV) – only R38 operates; the remaining relays are restrained.}
\label{fig6}
\end{figure}

\begin{figure}[!htb]
\centering
\hspace*{-0.8cm}
\includegraphics[scale=1.1]{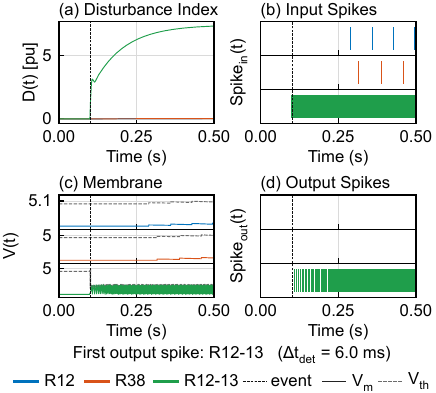}
\caption{Temporal signatures for an ABC fault on Line 12-13 (CIGRE MV) – only R12-13 operates; R12 and R38 are restrained.}
\label{fig7}
\end{figure}

\begin{figure*}[t]
  \centering
  \includegraphics[width=\textwidth]{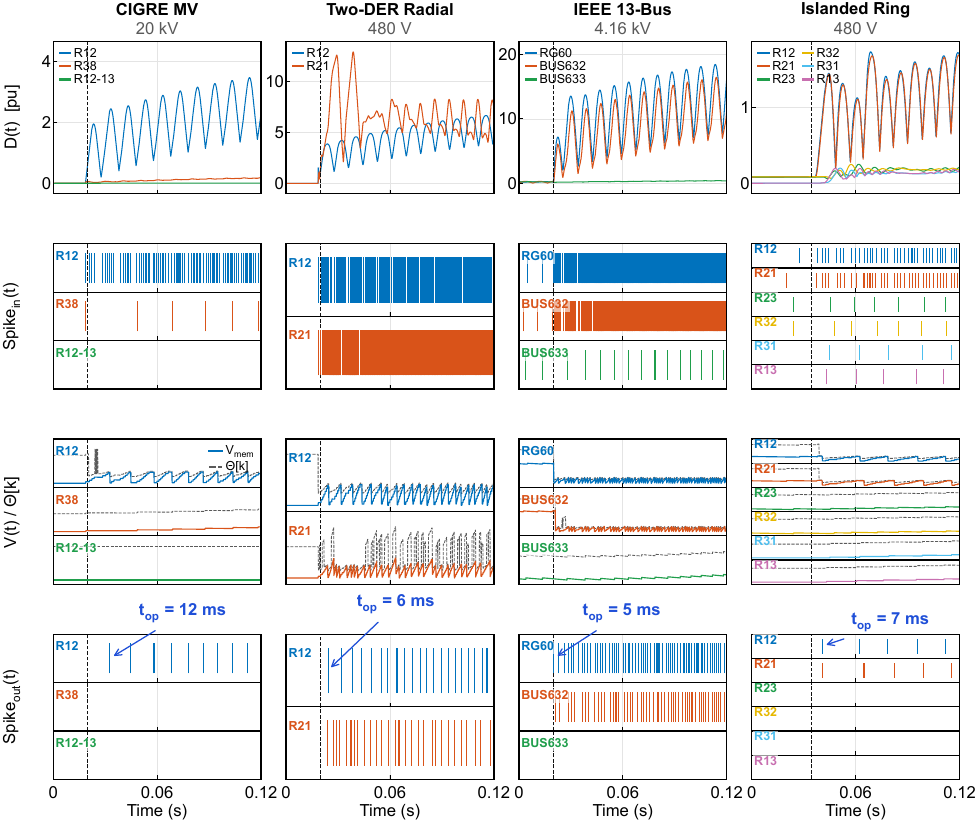}
  \caption{Scale-invariant transferability across four inverter-dominated networks (480~V--20~kV) without parameter retuning: Despite order-of-magnitude differences in the disturbance index $D(t)$, latency encoding ensures that only in-zone relays cross the adaptive threshold $\Theta[k]$ and generate trip spikes. The operating time remains within approximately one cycle ($\approx 5$--$13$~ms) across all voltage levels, demonstrating the preservation of selectivity and response speed across different electrical scales.}
  \label{fig8}
\end{figure*}

\subsection{Protection Performance in the CIGRE Medium-Voltage Benchmark}

The CIGRE medium-voltage (MV) benchmark is used to evaluate the proposed framework in a meshed distribution network with multiple fault-current paths. Three internal faults are considered: an A-phase-to-ground (AG) fault on Line~1--2, a BCG fault on Line~3--8, and an ABC fault on Line~12--13.

For the AG fault, Fig.~\ref{fig5} shows that only relay $R_{12}$ generates an output spike, with a first detection interval of 12.7~ms, while $R_{38}$ and $R_{12-13}$ remain restrained. Similarly, under the BCG fault, Fig.~\ref{fig6} shows selective operation of $R_{38}$ after 24.3~ms, with the remaining relays restrained. For the ABC fault, Fig.~\ref{fig7} demonstrates that $R_{12-13}$ generates the first output spike after 6.0~ms, while $R_{12}$ and $R_{38}$ remain restrained.

The detection times range from 6.0 to 24.3~ms across the three scenarios, reflecting the influence of fault characteristics and locations on the local electrical signatures and their evolution through neuronal integration. Importantly, only the relay associated with the faulted line responds in each case, while the other monitored relays remain restrained. This demonstrates that the proposed framework maintains directional selectivity despite the multiple fault-current paths of the meshed CIGRE network.

These results extend the evaluation beyond the two-DER microgrid and demonstrate the applicability of the common neuromorphic protection formulation to a larger meshed distribution system.

\subsection{Cross-Network Transferability and Scale-Invariant Protection Performance}

The cross-network transferability of the proposed framework is examined across four inverter-dominated networks with nominal voltage levels ranging from 480~V to 20~kV. Fig.~\ref{fig8} compares their disturbance indices and corresponding output-spike responses under a common neuromorphic decision formulation without parameter retuning. Despite substantial differences in disturbance magnitude arising from the different electrical ratings, asset-based normalization and latency encoding enable consistent neuronal processing across the evaluated networks.

As shown by the output-spike responses in Fig.~\ref{fig8}, only the designated in-zone relays generate trip spikes, while the remaining relays stay restrained. The representative operating times are approximately 12~ms for the CIGRE MV network, 6~ms for the two-DER radial microgrid, 5~ms for the IEEE 13-node feeder, and 7~ms for the islanded ring microgrid. Thus, protection responses remain within approximately one cycle (5--13~ms) across the evaluated voltage levels without network-specific parameter retuning.

These results demonstrate that the proposed framework preserves selective protection behaviour across substantially different electrical scales and network configurations. Importantly, this consistency results from the combined operation of asset-based normalization, latency encoding, neuronal integration, and adaptive threshold regulation, rather than normalization alone.

\begin{figure*}[t]
  \centering
  \includegraphics[width=\textwidth]{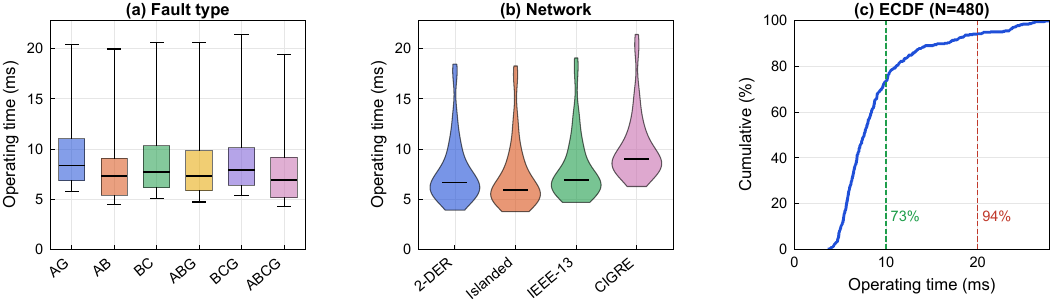}
  \caption{Operating-time characteristics (480 cases): (a) distribution by fault
type and (b) by network configuration, showing broadly similar response times
across fault types while network configuration---rather than fault
type---governs the spread, with the CIGRE MV network exhibiting the longest
operating times. (c) Cumulative distribution, indicating that 73\% of cases
operate within 10~ms and 94\% within one cycle (overall median 7.6~ms).}
\label{fig9}
\end{figure*}

\begin{figure*}[t]
  \centering
  \includegraphics[width=\textwidth]{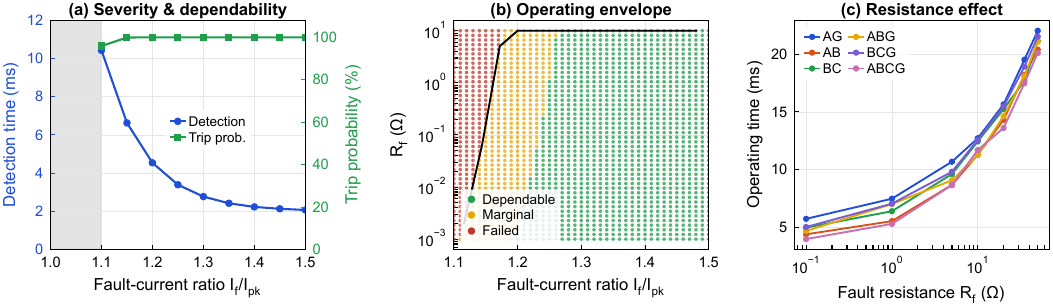}
  \caption{Sensitivity and operating envelope: (a) operating time decreases with
increasing fault-current severity (left axis), while the trip probability
approaches 100\% once the fault current exceeds the pickup threshold (right
axis; grey region indicates operation below pickup, where the relay does not
trip); (b) dependable, marginal, and failed operating regions in the ratio--$R_f$
plane; (c) operating time increases with fault resistance as the associated
disturbance signature becomes weaker.}
\label{fig10}
\end{figure*}

\begin{figure}[!t]
    \centering
    \includegraphics[width=0.95\columnwidth]{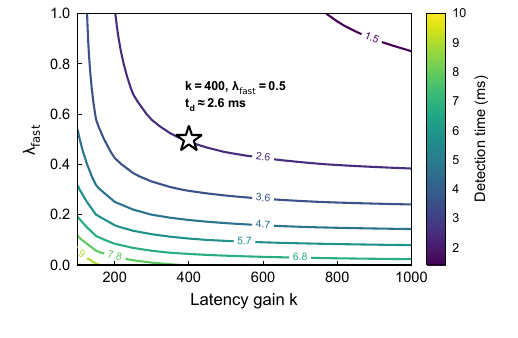}
    \caption{Sensitivity of detection time to the latency gain ($k$) and adaptive parameter ($\lambda_{\mathrm{fast}}$). The selected operating point, $k=400$ and $\lambda_{\mathrm{fast}}=0.5$, yields a detection time of approximately 2.6~ms.}
    \label{fig11}
\end{figure}

\begin{figure*}[t]
  \centering
  \includegraphics[width=0.9\textwidth]{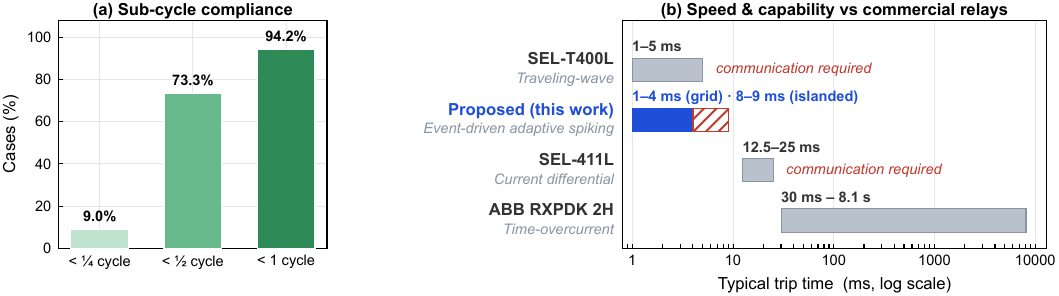}
  \caption{Protection speed and capability versus commercial relays: (a) Sub-cycle
compliance, expressed as the fraction of the 480 tested cases operating within
$\tfrac{1}{4}$, $\tfrac{1}{2}$, and one power cycle (9.0\%, 73.3\%, and 94.2\%,
respectively), (b) Typical trip times (logarithmic scale), together with the
operating principle and communication requirements of each protection scheme.
The proposed event-driven relay operates within 1--4 ms under grid-connected
conditions and 8--9~ms under islanded conditions, without requiring communication
links. In contrast, traveling-wave~\cite{selt400l} and differential~\cite{sel411l}
protection schemes require high-speed communication links, whereas conventional
overcurrent protection with time grading~\cite{abb_rxpdk2h} may exhibit operating
times extending to several seconds.}
  \label{fig12}
\end{figure*}

\section{Sensitivity Analysis}

\subsection{Operating-Time Characteristics Across Fault Types and Network Configurations}

Operating-time characteristics are evaluated across 480 cases covering four
network configurations. The distributions in Fig.~\ref{fig9}(a) and (b) show broadly
similar responses across fault types, while the CIGRE MV network exhibits longer
operating times than the other configurations, indicating that network
configuration, rather than fault type, primarily governs response-time
variation. The cumulative distribution in Fig.~\ref{fig9}(c) further shows that 73\% of
cases operate within 10~ms and 94\% within one power cycle, with an overall
median of 7.6~ms.

\subsection{Protection Sensitivity and Operating Envelope}

Protection sensitivity is evaluated under varying fault-current severity and resistance. As shown in Fig.~\ref{fig10}(a), operating time decreases with increasing fault-current severity, while trip probability approaches 100\% above the pickup threshold. The operating regions in Fig.~\ref{fig10}(b) distinguish dependable, marginal, and failed detection in the ratio--$R_f$ plane. In Fig.~\ref{fig10}(c), operating time increases with fault resistance due to the weakening of the associated disturbance signature. These results characterize the electrical conditions governing protection sensitivity and dependability.

The effects of latency gain $k$ and adaptive parameter $\lambda_{\mathrm{fast}}$ on detection time are examined in Fig.~\ref{fig11}. The detection-time surface illustrates their combined influence, with the selected operating point ($k=400$, $\lambda_{\mathrm{fast}}=0.5$) yielding approximately $2.6~\mathrm{ms}$. This parameter combination is adopted for the subsequent cross-network evaluation.

\subsection{Protection Speed and Comparison with Existing Relays}

The sub-cycle response distribution in Fig.~\ref{fig12}(a) shows that 9.0\%, 73.3\%, and 94.2\% of the 480 cases operate within one-quarter, one-half, and one power cycle, respectively. The comparison in Fig.~\ref{fig12}(b) further shows that the proposed framework achieves operating times of approximately 1--4~ms under grid-connected conditions and 8--9~ms under islanded conditions, without communication links. In contrast, traveling-wave and differential schemes require high-speed communication links, while time-graded overcurrent protection can exhibit substantially longer operating times. These results highlight the potential of the proposed framework for fast, communication-free protection across heterogeneous networks.

\section{Conclusions}

This paper proposed a scale-invariant neuromorphic protection framework integrating asset-based normalization, multi-domain disturbance representation, event-driven spike encoding, leaky integrate-and-fire dynamics, adaptive threshold regulation, and directional supervision within a unified mathematical formulation. Evaluations across four heterogeneous network configurations demonstrated selective fault detection without network-specific parameter retuning. Across 480 cases, the framework achieved an overall median operating time of $7.6~\mathrm{ms}$, with 94.2\% of cases operating within one power cycle. It also maintained restraint during the tested load disturbance and provided communication-free protection under both grid-connected and islanded conditions.

The results establish the cross-network transferability of the proposed framework across different network topologies, operating conditions, and electrical characteristics. Future work will extend the theoretical characterization of scale invariance to the complete protection chain, establish operating boundaries under weak-source and high-resistance faults, and investigate scalability through real-time hardware validation on larger distribution networks.





\footnotesize{\bibliographystyle{IEEEtran}}
\bstctlcite{BSTctl}

\bibliography{ref}

@article{hooshyar2017microgrid,
  title={Microgrid protection},
  author={Hooshyar, Ali and Iravani, Reza},
  journal={Proceedings of the IEEE},
  volume={105},
  number={7},
  pages={1332--1353},
  year={2017},
  publisher={IEEE}
}

@article{manson2021practical,
  title={Practical microgrid protection solutions: Promises and challenges},
  author={Manson, Scott and McCullough, Ed},
  journal={IEEE Power and Energy Magazine},
  volume={19},
  number={3},
  pages={58--69},
  year={2021},
  publisher={IEEE}
}

@article{aliyu2025comprehensive,
  title={Comprehensive review of fault ride-through techniques for grid-forming inverters: current approaches, challenges, and future directions},
  author={Aliyu, Aminu Jibrin and Kannan, Ramani},
  journal={Engineering Research Express},
  volume={7},
  number={2},
  pages={022302},
  year={2025},
  publisher={IOP Publishing}
}

@article{paladhi2024reliable,
  title={A reliable accelerated protection scheme for converter-dominated power networks},
  author={Paladhi, Subhadeep and Hong, Qiteng and Booth, Campbell},
  journal={IEEE Transactions on Industry Applications},
  volume={60},
  number={4},
  pages={5599--5608},
  year={2024},
  publisher={IEEE}
}

@article{roy2019towards,
  title={Towards spike-based machine intelligence with neuromorphic computing},
  author={Roy, Kaushik and Jaiswal, Akhilesh and Panda, Priyadarshini},
  journal={Nature},
  volume={575},
  number={7784},
  pages={607--617},
  year={2019},
  publisher={Nature Publishing Group UK, London}
}

@article{tavanaei2019deep,
  title={Deep learning in spiking neural networks},
  author={Tavanaei, Amirhossein and Ghodrati, Masoud and Kheradpisheh, Saeed Reza and Masquelier, Timoth{\'e}e and Maida, Anthony},
  journal={Neural networks},
  volume={111},
  pages={47--63},
  year={2019},
  publisher={Elsevier}
}

@article{kamoona2026robust,
  title={Robust Industrial Cyber Physical Classification Using Neuromorphic Temporal Embeddings and Hybrid SNN XGBoost Under Machine Unlearning Attacks},
  author={Kamoona, Ammar and Koushkbaghi, Sajad and Jalili, Mahdi and McTaggart, Peter and Yu, Xinghuo},
  journal={arXiv preprint arXiv:2609.09564},
  year={2026}
}

@article{barranco2023adaptive,
  title={Adaptive protection for active distribution networks: An approach based on fuses and relays with multiple setting groups},
  author={Barranco-Carlos, Adri{\'a}n and Orozco-Henao, Cesar and Mar{\'\i}n-Quintero, Juan and Mora-Fl{\'o}rez, Juan and Herrera-Orozco, Andres},
  journal={IEEE Access},
  volume={11},
  pages={31075--31091},
  year={2023},
  publisher={IEEE}
}

@article{narang2021overview,
  title={An overview of issues related to IEEE Std 1547-2018 requirements regarding voltage and reactive power control},
  author={Narang, David and Mahmud, Rasel and Ingram, Michael and Hoke, Anderson},
  year={2021},
  publisher={National Renewable Energy Laboratory (NREL), Golden, CO (United States)}
}

@article{prabhakar2026synapse,
  title={Synapse-inspired energy networks: a neuromorphic approach to microgrid protection without communication links},
  author={Prabhakar, Saurabh and Panigrahi, Bijaya Ketan and Blaabjerg, Frede and Sahoo, Subham},
  journal={Communications Engineering},
  volume={5},
  number={1},
  pages={90},
  year={2026},
  publisher={Nature Publishing Group UK, London}
}

@article{sahoo2025spike,
  title={Spike talk: Genesis and neural coding scheme translations},
  author={Sahoo, Subham},
  journal={IEEE Transactions on Smart Grid},
  volume={16},
  number={3},
  pages={2659--2670},
  year={2025},
  publisher={IEEE}
}

@article{diao2024neuromorphic,
  title={Neuromorphic event-driven semantic communication in microgrids},
  author={Diao, Xiaoguang and Song, Yubo and Sahoo, Subham and Li, Yuan},
  journal={IEEE Transactions on Smart Grid},
  volume={15},
  number={5},
  pages={4300--4314},
  year={2024},
  publisher={IEEE}
}

@article{samkari2023time,
  title={Time-domain protection scheme for microgrids with aggregated inverter-based distributed energy resources},
  author={Samkari, Husam S and Johnson, Brian K},
  journal={IEEE access},
  volume={11},
  pages={13232--13242},
  year={2023},
  publisher={IEEE}
}

@article{quispe2022transmission,
  title={Transmission line protection challenges influenced by inverter-based resources: A review},
  author={Quispe, Juan Carlos and Ordu{\~n}a, Eduardo},
  journal={Protection and Control of Modern Power Systems},
  volume={7},
  number={1},
  pages={28},
  year={2022},
  publisher={Springer}
}

@article{liu2022transient,
  title={Transient wavelet energy-based protection scheme for inverter-dominated microgrid},
  author={Liu, Di and Dy{\'s}ko, Adam and Hong, Qiteng and Tzelepis, Dimitrios and Booth, Campbell D},
  journal={IEEE Transactions on Smart Grid},
  volume={13},
  number={4},
  pages={2533--2546},
  year={2022},
  publisher={IEEE}
}

@article{zaben2024machine,
  title={Machine learning methods for fault diagnosis in AC microgrids: A systematic review},
  author={Zaben, Muiz M and Worku, Muhammed Y and Hassan, Mohamed A and Abido, Mohammad A},
  journal={IEEE access},
  volume={12},
  pages={20260--20298},
  year={2024},
  publisher={IEEE}
}

@article{amiri2020integrated,
  title={Integrated protection scheme for both operation modes of microgrid using S-Transform},
  author={Amiri, E Maali and Vahidi, B},
  journal={International Journal of Electrical Power \& Energy Systems},
  volume={121},
  pages={106051},
  year={2020},
  publisher={Elsevier}
}

@article{dua2022microgrid,
  title={Microgrid differential protection based on superimposed current angle employing synchrophasors},
  author={Dua, Gagandeep Singh and Tyagi, Barjeev and Kumar, Vishal},
  journal={IEEE Transactions on Industrial Informatics},
  volume={19},
  number={8},
  pages={8775--8783},
  year={2022},
  publisher={IEEE}
}

@inproceedings{karabulut2025generalization,
  title={Generalization of graph neural network models for distribution grid fault detection},
  author={Karabulut, Burak and Manna, Carlo and Develder, Chris},
  booktitle={2025 IEEE International Conference on Communications, Control, and Computing Technologies for Smart Grids (SmartGridComm)},
  pages={1--7},
  year={2025},
  organization={IEEE}
}

@IEEEtranBSTCTL{BSTctl,
  CTLdash_repeated_names = "no"
}

@misc{abb_rxpdk2h,
  author       = {{ABB (Hitachi Energy)}},
  title        = {Directional Overcurrent Relay Type {RXPDK~2H} and {RAPDK}},
  howpublished = {Product Guide and Technical Manual, Doc.\ 1MRK509007-BEN, Rev.\ C},
  note         = {[Online]. Available: \url{https://publisher.hitachienergy.com/download?DocumentID=1MRK509007-BEN&LanguageCode=en&Action=download}}
}

@misc{sel411l,
  author       = {{Schweitzer Engineering Laboratories}},
  title        = {{SEL-411L Advanced Line Differential and Distance Relay}},
  howpublished = {Product Data Sheet, PDS411L-01},
  year         = {2018},
  note         = {[Online]. Available: \url{https://www.pesin.com.ph/sites/default/files/product-file/411l_ds_20180329.pdf}}
}

@misc{selt400l,
  author       = {{Schweitzer Engineering Laboratories (SEL)}},
  title        = {{SEL-T400L} Time-Domain Line Protection},
  howpublished = {High-Speed Protection Manual, PDST400L-01},
  year         = {2017},
  note         = {[Online]. Available: \url{https://www.scribd.com/document/921352180/SEL-T400}}
}

\end{document}